\documentclass[conference]{IEEEtran}

\usepackage{amsmath, amssymb, bbm}
\usepackage{steinmetz,mathcomSTEv4}
\usepackage{xcolor}
\usepackage{cite}

\newcommand{\rr}{\mathrm {r}}

\newcommand{\ntq}{k}
\newcommand{\bps}{\rm bps}

\newcommand{\fact}[2]{\mathop{\left(\!\!\!\begin{array}{c} {#1}\\{#2} \end{array}\!\!\!\right)}\nolimits}
\newtheorem{remark}{Remark}

\usepackage{tikz}
\usetikzlibrary{positioning}
\usetikzlibrary{arrows,fit}
\usetikzlibrary{shapes.geometric}
\usetikzlibrary{positioning,calc}
\usetikzlibrary{decorations.markings,shapes,arrows}

\usepackage{epstopdf}
\usepackage{pgfplots,setspace}
\usepackage{circuitikz}

\usepackage{algorithm}
\usepackage{algorithmic}

\tikzstyle{int}=[draw, fill=blue!10, minimum height = .5 cm, minimum width=1 cm,thick ]
\tikzstyle{int1}=[draw,  minimum height = .15 cm, minimum width=1 cm,thick ]
\tikzstyle{tri}=[isosceles triangle, draw=black, fill=blue!10, minimum height = .1 cm, thin ]
\tikzstyle{sum}=[circle, fill=blue!10, draw=black ]	

\title{
Comparison-limited Vector Quantization
}

\author{
	\IEEEauthorblockN{Joseph Chataignon }
\IEEEauthorblockA{
T\'el\'ecom Saint-\'Etienne \\
Universit\'e Jean Monnet,  France\\
	\texttt{joseph.chataignon@telecom-st-etienne.fr}
}
	\and
	\IEEEauthorblockN{Stefano Rini}
	\IEEEauthorblockA{
		Department of Electrical and Computer Engineering\\
		National Chiao Tung University, Taiwan\\
		\texttt{stefano@nctu.edu.tw}
	}
}

\begin{document}
	
		\maketitle
	
\begin{abstract}
%
A variation of the classic vector quantization problem is considered, in which the analog-to-digital (A2D) conversion is not constrained by the cardinality of the output but rather by the number of comparators available for quantization. 
%
%
%
More specifically, we consider the scenario in which a vector quantizer of dimension $d$ is comprised of $\ntq$ comparators, each receiving a linear combination of the inputs and producing zero/one when this signal is above/below a threshold. 
Given a distribution of the inputs and a distortion criterion, the value of the linear combinations and thresholds are to be configured so as to minimize the distortion between the quantizer input and its reconstruction.
This vector quantizer architecture naturally arises in many A2D conversion scenarios in which the quantizer's cost and energy consumption are severely restricted.
%
%
%
%
%
For this novel vector quantizer architecture, we propose an algorithm to determine the optimal configuration and 
provide the first performance evaluation for the case of uniform and Gaussian sources.
\end{abstract}

\section{Introduction}
\begin{figure}
	\centering
	\includegraphics[width=1.1\linewidth]{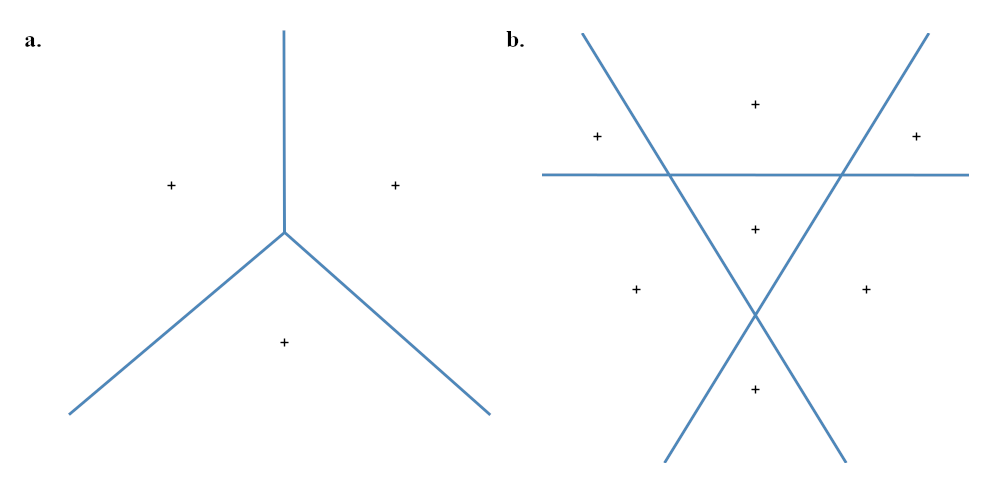}
	\caption{Schematic representation of: a. a 2-dimensional quantizer with rate $R=1.5 \ \bps$
	    and b. a 2-dimensional quantizer utilizing $t=3$ comparators.}
	\label{fig:quantization}
\end{figure}

Quantization, that is transforming continuous amplitude values into discrete ones so as to minimize a prescribed distortion measure 
subject to a output cardinality constraint,  is one of the fundamental signal processing operations.
%
%
%
%
As such, quantization has been studied in a number of contexts and a vast amount of results have been derived for this problem.
In this paper we consider a variation 
of this problem, so far neglected in the literature,  that is 
relevant in the design of low-cost, energy-efficient quantizers.
%
%
Vector quantizers are typically manufactured using op-amp comparators that obtain a linear combinations of the quantizer inputs and a bias and produce a zero/one voltage whether the comparison between these two signals is positive/negative.
Generally speaking,  comparators are components with high power consumption and manufacturing cost: for this reason it is reasonable to evaluate the cost of a quantizer in terms of the number of comparators it requires.
%
Let us consider the case of a two dimensional quantizer ($d=2$) in which each dimension is quantized with a rate of 1.5 bits-per-sample ($R=1.5 \ \bps$).
When quantizing a generic source, the three discretization points are separated in Voronoi regions as in Fig. \ref{fig:quantization}.a so that the number of comparators required is $\ntq=3$.
More generally, since every two reconstruction points are separated by an edge of the Voronoi region, a vector quantizer require $2^R(2^R-1)/2$ comparators.
This scaling of the  quantizer cost is generally valid for low-rate quantizer, since in this scenario the number of neighbors of each reconstruction point is large.
Given that this quantization regime is naturally associated with low cost devices,  the cost scaling seems to be particularly disadvantageous.
%
%
A natural question that naturally arises is whether a better scaling of the quantizer cost can be attained. 
%
To address this question, note  that number of comparators $\ntq$ equals the total number of hyperplane segments in the Voronoi regions.
%
Accordingly, the best scaling is attained when the $\ntq$ hyperplanes induce the largest number of  partitions of the space of dimension $d$.
%
%
After some geometrical consideration, one realizes that for the case of $d=2$ and $R=1.5$, the largest number of partition is indeed $7$ and corresponds to the configuration in Fig. \ref{fig:quantization}.b.
It is now apparent that there exists a large gap in the optimal quantizer design whether one considers a constraint on the number of points used in reconstruction or the number of comparators employed by the quantizer

%
%
%

\subsubsection*{Relevant Results}
The idea of accounting for the number of comparators required for A2D conversion emerges from the work in \cite{rini2017general}: here the authors investigate the capacity of a MIMO channel with output quantization constraint, i.e. the MIMO channel in which the channel output is processed at the receiver using a finite number of one-bit threshold quantizers. 
The channel model of \cite{rini2017general} is relevant in mm-wave communications which allows for a large number of receive antennas, while the number of A2D conversion modules remains small due to limitation in the energy and costs of RF modules.
Building on an idea of \cite{mo2014high}, a connection between combinatorial geometry and the MIMO channel with output quantization constraints of \cite{rini2017general} is drawn in \cite{khalili2018mimo}. In particular, in \cite{khalili2018mimo} it is shown that each quantizer can be interpreted as an hyperplane bisecting the transmitter signal space; for this reason  the largest rates are attained by the configuration allowing for the largest number of partitions induced by the set of hyperplanes.
To the best of our knowledge, the problem of hardware limited quantization has so far only being considered in \cite{shlezinger2018hardware}. 
The difference between our approach and that of  \cite{shlezinger2018hardware} is in focusing on vector quantization and considering the comparator limitations on the vector, rather than scalar, quantizer input.
%
%
%
%

\subsubsection*{Contribution}
In the following, we define comparison limited vector quantization  problem in its full generality as a variation of the classical vector quantization problem. 
We also provide a first algorithm for the quantizer design: although not optimal, this first approach investigates the combinatorial geometric aspects of the optimal quantization design problem. 
Numerical evaluations are provided for the case of an iid Gaussian and uniform source.  
The performance of the proposed quantizer is compared with the classic Max-Lloyd quantizer design \cite{max1960quantizing,lloyd1982least}.

\subsubsection*{Paper Organization}
The paper is organized as follows: Sec. \ref{sec:Vector Quantizer Model} presents the vector quantization model,  Sec. \ref{sec:Design Algorithm} introduces the proposed design algorithm. 
Sec. \ref{sec:Simulation results} provides relevant numerical evaluations. 
Finally, Sec. \ref{sec:Conclusion} concludes the paper.

\subsubsection*{Notation}
In the remainder of the paper, all logarithms are taken in base two.
With $\xv= [x_1, \ldots, x_N] \subseteq \Xcal^n$ we indicate a sequence of elements from $\Xcal$ with length $N$.
The notation $\xv_i^j$ indicates the substring $[x_i, \ldots, x_j]$ of $\xv$.
The function $\sign(\xv)$ returns a vector with values in $\{-1,+1\}$ which equals the sign of each entry of the vector $\xv$.
Random vector  are indicated as {\small $\Uv=[U_{1} \ldots U_{L}]^T \in \Rbb^L$}. 
%
%
%
The set $\{1,\ldots, N \}$ is indicated as $[N]$.

\section{Comparison-limited vector quantizer model}
\label{sec:Vector Quantizer Model}

\begin{figure*}
	\centering
	\resizebox{1.05\textwidth}{!}{
		\begin{tikzpicture}[node distance=2.5cm,auto,>=latex]
		\tikzset{point/.style={coordinate},
			block/.style ={draw, thick, rectangle, minimum height=4em, minimum width=6em},
			line/.style ={draw, very thick,-},
		}
		\node (b1) at (0,0){};
		\node (b2) [block, minimum height = 1 cm, minimum width=1cm,thick =1cm] at (2,0) {\small $d$-parser};
		%
		\draw (b1) [->,line width=0.5 pt]-- node[above]{\small $X_i$} (b2);

		\node (d1) at (6,1.5)  [block,minimum height = 0.5 cm,minimum width=.5cm] {\small  $\vv_1$};
		\node (d2) at (6,0.5)  [block,minimum height = 0.5 cm,minimum width=.5cm] {\small  $\vv_{2}$};
		\node (d3) at (6,-0.5)  [block,minimum height = 0.5 cm,minimum width=.5cm] {\small  $\vv_{3}$};
		\node (d4) at (6,-1.5)  [block,minimum height = 0.5 cm,minimum width=.5cm] {\small  $\vv_{4}$};

		\node (b3) at (5.5,0){};
		
		\draw (b2) [line width=0.5 pt]-- node[above]{\small $\Xv_n=X_{di+1}^{d(i+1)}$} (b3);
		
		\draw (b3.west) [line width=0.5 pt]|- (d1);
		\draw (b3.west) [line width=0.5 pt]|- (d2);
		\draw (b3.west) [line width=0.5 pt]|- (d3);
		\draw (b3.west) [line width=0.5 pt]|- (d4);
		
		\node  at (7,-1.5) (e11) [sum,scale=0.35]{$+$};
		\node  at (7,-.5) (e21) [sum,scale=0.35]{$+$};
		\node  at (7,+.5) (e31) [sum,scale=0.35]{$+$};
		\node  at (7,+1.5) (e41) [sum,scale=0.35]{$+$};

		\node  at (7-0.4,-1.5-0.4) (th1) {\scriptsize   $t_4$};
		\node  at (7-0.4,-.5-0.4) (th2) {\scriptsize   $t_3$};
		\node  at (7-0.4,+.5-0.4) (th3) {\scriptsize   $t_2$};
		\node  at (7-0.4,+1.5-0.4) (th4) {\scriptsize   $t_1$};
		
		\node  at (8.25,-1.5) (quant1) [int1] {\small  $\sign(\cdot)$};
		\node  at (8.25,-.5) (quant2)  [int1] {\small  $\sign(\cdot)$};
		\node  at (8.25,+.5) (quant3)  [int1] {\small  $\sign(\cdot)$};
		\node  at (8.25,+1.5) (quant4) [int1] {\small  $\sign(\cdot)$};
		
		\draw (th1) [->]-| (e11);
		\draw (th2) [->]-| (e21);		
		\draw (th3) [->]-| (e31);
		\draw (th4) [->]-| (e41);
		
		\draw (d1) -- (e41);
		\draw (d2) -- (e31);		
		\draw (d3) -- (e21);
		\draw (d4) -- (e11);
		
		\draw (quant4) -- (e41);
		\draw (quant3) -- (e31);		
		\draw (quant2) -- (e21);
		\draw (quant1) -- (e11);

		\node (d5) at (11,0)  [block,minimum height = 0.5 cm,minimum width=.5cm] {\small $n$-parser};
		
		\draw (quant4)-| (d5.west);
		\draw (quant3)-| (d5.west);
		\draw (quant2)-| (d5.west);		
		\draw (quant1)-| (d5.west);								
		
		\node  at (9.5,1.5+0.35)  {\small  $Y_{1n}$};
		\node  at (9.5,.5+0.35)  {\small  $Y_{2n}$};
		\node  at (9.5,-.5+0.35)  {\small  $Y_{3n}$};
		\node  at (9.5,-1.5+0.35)  {\small  $Y_{4n}$};
		
		\node (d6) at (15,0)  [block,minimum height = 0.5 cm,minimum width=.5cm, text width=1.5cm] {source encoder};
		\draw (d5) [->,line width=0.5 pt]-- node[above]{\small $\Yv_n=Y_{ni+1}^{n(i+1)}$} (d6);
		
		\node (d7) at (19,0)  [block,minimum height = 0.5 cm,minimum width=.5cm, text width=1.5cm] {source decoder};
		
		\draw (d6) [->,line width=0.5 pt]-- node[above]{\small $m_n \in [2^{dR}]$} (d7);
		
		\node (d8) at (21,0) {};
		
		\draw (d7) [->,line width=0.5 pt]-- node[above]{\small $\Xhv_n$} (d8);

		\end{tikzpicture}
	}
	\caption{Comparison-limited quantizer with linear pre-processing for $\ntq=4$ and rate $R$}
	\label{fig:system model}
\end{figure*}
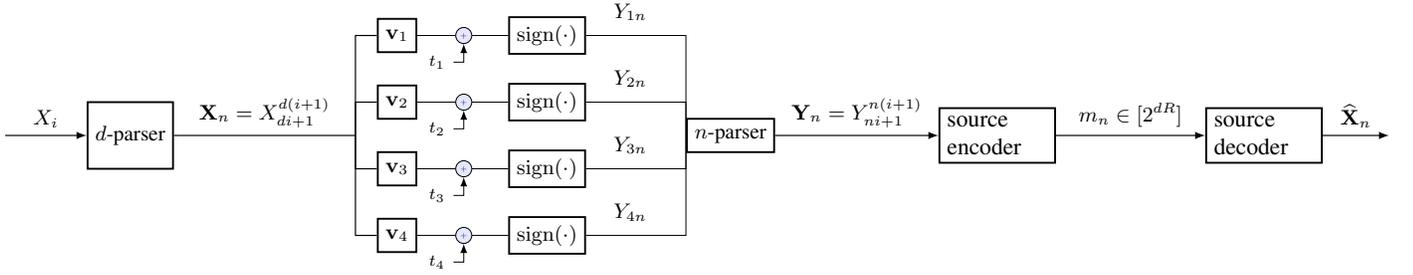

We consider the source quantization scenario in Fig.  \ref{fig:system model}: the source sequence $\{X_i\}_{i\in \Nbb}$  with support $\Xcal$.
%
The source sequence is parsed in super-symbols $\{\Xv_n\}_{n\in \Nbb}$ of dimension $d$ with $\Xv_n=[X_{dn+1}, \ldots, X_{d(n+1)}]$ where $d$ is referred to as the \emph{dimension} of the vector quantizer.
The $j^{\rm th}$ comparator obtains a linear combination of the each super-symbol $\Xv_n$ and produces the signal $Y_{jn}$  as
\ea{
Y_{jn} = \sign\lb \vv_j \Xv_n+t_j \rb,
\label{eq:Yjn}
}
for $j \in [\ntq]$. The value $\ntq$ is is referred to as the \emph{resolution} of the vector quantizer; $\vv_j \in \Rbb^{d}$, $t_j \in\Rbb$ are fixed and known. 
The outputs of the $\ntq$ quantizers in \eqref{eq:Yjn} is more conveniently expressed in vector for as
\ea{
\Yv_n=\sign \lb \Vv \Xv_n+\tv \rb,
}
where $\Vv \in \Rbb^{\ntq \times d}$ is such that the $i^{\rm th}$ row corresponds to the vector $\vv_i$ in in \eqref{eq:Yjn}. 
Similarly, $\tv \in \Rbb^{\ntq}$ has the  $i^{\rm th}$  entry equal to $t_i$ and  $\Yv_n=[Y_{1n},\ldots,Y_{\ntq n}]$.
The set $[\Vv,\tv]$ is referred to as the configuration of the linear combiner.
%
%
The supersymbol $\Yv_n$ is provided to a source encoder 
%
%
that produces a bit-restricted representation of the quantizers' output as $m_n \in [2^{\lfloor dR \rfloor}]$ where $R$ referred to as  the \emph{rate} of the quantizer through the \emph{source encoding mapping}
\ea{
f_{\rm enc}: \ \{-1,+1\}^{\ntq} \goes [2^{\lfloor dR \rfloor}].
}
The message $m_n$ is provided to a source decoder which produces a reconstruction of the source super-symbol $\Xv_n$, $\Xhv_n= [\Xh_{dn+1}, \ldots, \Xh_{d(n+1)}]$ 
with $ \Xh_i \in \widehat{\Xcal}$ ,  thorough the \emph{source decoding mapping}
\ea{
f_{\rm dec}: \ [2^{\lfloor dR \rfloor}] \goes \widehat{\Xcal}^d.	
}
The performance of the vector quantizer is evaluated through a distortion measure 
\ea{
\rho^n(X^n;\Xh^n): \ \Xcal\times \widehat{\Xcal} \goes \Rbb^+,
} 
for  $n	\in \Nbb$ which is assumed non-decreasing in $n$.
For a given configuration of the linear combiners $[\Vv,\tv]$, source encoder/decoder mappings $f_{\rm enc}$/$f_{\rm dec}$,  and 
given a distortion measure between input and reconstruction sequence $\rho^n(X^n;\Xh^n)$,  the performance of the quantizer is evaluated as
\ea{
\overline{\rho}= \limsup_{n \goes \infty} \f 1 n \rho^n(X^n;\Xh^n).
}
The optimal quantizer performance for the distortion $\overline{\rho}$, dimension $d$, resolution $\ntq$ and rate $R$ is obtained as 
\ea{
D(d,\ntq,R)=\inf \overline{\rho},
\label{eq:opt}
}
where the infimization is over all linear combiner configurations and source encoder/decoder mappings.

\subsubsection*{Comparison-limited distortion-rate function}
%
Let $\ntq=2^{\al}$,  then the comparison-limited  distortion-rate function is defined as 
\ea{
D(R,\al)=\lim_{d \goes \infty, \ntq=e^{\al d} }D(d,\ntq,R),
\label{eq: D R al}
}
for $D(d,\ntq,R)$ in \eqref{eq:opt}, that is $D(\al,R)$ is the minimum distortion attainable as the quantizer dimension grows to infinity while the message support grows as $2^{dR}$ and the number of quantizers as $2^{\al d}$.
One can also show that $D(R,\al)$  in \eqref{eq: D R al} for $\al\geq 2R$ correspond to the classical rate distortion function \cite[Ch. 13]{cover2012elements}.

%
%

\begin{remark}
The above vector quantizer architecture formulation is rather general: in the remainder we consider only the case in which $R=\infty$ in  \eqref{eq:opt}, i.e.  $D(d,\ntq,\infty)$.
This corresponds to the scenario in which the communication rate between the source encoder and the decoder in \eqref{fig:system model} unbounded.
\end{remark}

\subsubsection*{Some combinatorial notions}
\label{Combinatorial Interlude}
%
%
In the following, we utilize some simple combinatorial concepts which we briefly introduce here. 

A  \emph{hyperplane arrangement}  $\Acal$ is a finite set of $n$ affine hyperplanes in $\Rbb^m$ for some $n,m \in \Nbb$.
A hyperplane arrangement  $\Acal=\{ \xv \in \Rbb^m, \ \av_i^T \xv = b_i \}_{i=1}^n$ can be expressed as $\Acal=\lcb \xv,  \  \Av \xv=\bv\rcb$
where $\Av$ is obtained by letting each row $i$ correspond to $\av_i^T$ and defining $\bv=[b_1 \ldots b_n]^T$.
A  plane arrangement is said to be in General Position (GP) if and only if every $n \times n$ sub-matrix of  $\Av$ has non zero determinant \cite{Tcover1965Geometrical}.
%
An hyperplane arrangement induces a partition of the space $\Rbb^m$ in a number of regions.
\begin{lem}
	\label{lem:hyperplane}
	A hyperplane arrangement of size $n$  in  $\Rbb^m$ divides $\Rbb^m$ into at most 
		\begin{equation}
		\rr(m,n)=
		\sum_{i=0}^{m} \fact{n}{i} \leq 2^n,
		\end{equation}
	regions.
	Hyperplanes in GP divide the space in $\rr(m,n)$ regions.
\end{lem}
%
We see from Lem. \ref{lem:hyperplane} that the largest number of reconstruction points for $D(d,\ntq,\infty)$ is $\rr(d,\ntq)$. 

\section{Design algorithm}
\label{sec:Design Algorithm}

In the following section, we propose an algorithm to numerically determine the optimal linear combiner configuration and source reconstruction 
attaining $D(d,\ntq,\infty)$.
For simplicity we assume the case of  iid sources  to be reconstructed under  Mean Squared Error (MSE) distortion, i.e. 
\ea{
\rho^n(X^n;\Xh^n)=\f 1 n \sum_{i=1}^n |X_i-\Xh_i|^2.
\label{eq:mse}
} 
In this scenario,  one can show that \eqref{eq:opt} simplifies as
\ea{
D(d,k,R)=
\sum_{i=1}^d \Ebb \lsb \|X_i -\Xh_i \|^2 \rsb,
\label{eq:simply to expect}
}
where $\Xv=[X,\ldots,X]$ and $\Xhv=[\Xh_1,\ldots,\Xh_d]$ for $X,\Xh_i $ in \eqref{eq:simply to expect} are iid distributed, so that the subscript $n$ can be dropped.
\footnote{That is, every super-symbol $\Xv_i$ is quantized in the same manner, regardless of $n$.}

Similarly to the classic Max-Lloyd algorithm, the optimal quantizer design for the model in Sec. \ref{sec:Vector Quantizer Model} can be divided into two optimization steps to be iterated until convergence:
(i) the optimization  of the set of reconstruction points {\footnotesize $\Xhv$} for a given combiner configuration $[\Vv \ \tv]$ and (ii) the optimization of the combiner configuration for a given set of  set of reconstruction points.
The optimization step (i) is rather straightforward as the reconstruction points are chosen as the centroids of the regions induced by the hyperplane arrangement $[\Vv \ \tv]$ in $\Rbb^d$ \cite{orlik2013arrangements}.
The optimization step (ii) is rather more involved and  we propose two methods for this optimization:
a \emph{global} configuration update and a \emph{local} one as described in Algorithm \ref{al:one}. 
In the global configuration update, all hyperplanes are randomly perturbed with a perturbation of variance decreasing with the iteration number.
%
%
%
In the local configuration update, a hyperplane is selected at random and its position is optimally determined so as to minimize the MSE of the reconstruction points separated by such hyperplane.
One of these two methods is selected at random at every iteration, with the probability to use the global configuration update exponentially decreasing over iterations.

The reasoning between local and global update is as follows: consider a lower triangular matrix $M$ of size $\rr(d,\ntq) \times \rr(d,\ntq)$ and let the element in position  $i\times j$ in $M$ equal to the index of an hyperplane separating $\Xh_i$ and $\Xh_j$ if such reconstruction point exists and zero otherwise. \footnote{Note that some hyperplane arrangements induce less that $\rr(d,\ntq)$ regions: we assume that there exists a natural numbering of the possible $\rr(d,\ntq)$ regions.}
The matrix $M$ can be though of as one among a finite number of ways in which hyperplanes separate the reconstruction points.
In this view, the local update maximizes the quantizer performance in a given value of $M$. 
The global update, instead, allows ``hyperplanes to jump over centroids'', resulting in a different  matrix $M$.

\begin{algorithm}
\caption{Optimization step (ii) in Sec. \ref{sec:Design Algorithm}.}
\begin{algorithmic}
\STATE generate a random starting configuration
\FOR{$k$ = 1 to $totalIterations$}
\IF{$random(0,1) < \exp(-0.1 k)$}
\STATE $directions \leftarrow$ generate random configurations within a certain distance of the original one
\ELSE
\STATE $directions \leftarrow$ generate new configurations by changing randomly one of the variables of one of the hyperplanes
\ENDIF
\STATE compute MSE for each direction generated
\STATE choose the direction with the lowest MSE
\ENDFOR
\end{algorithmic}
\label{al:one}
\end{algorithm}

A crucial step in step (ii) of the design algorithm is the evaluation of the MSE for a given hyperplane configuration and reconstruction points. 
Given numerical precision limitations, the MSE evaluation has to be approximated using numerical  integration methods and particle filters as in Algorithm \ref{al:two}.
More specifically, random points are generated and assigned to the corresponding reconstruction point until a minimum number of points per region is attained.
%


\begin{algorithm}
\caption{MSE estimation in Sec. \ref{sec:Design Algorithm}.}
\begin{algorithmic}
\STATE $error \leftarrow 0$
\STATE $regions$ , $centroids$ , $pointsPerRegion$ = empty arrays
\FOR{$numberOfPointsCentroids$ times}
\STATE $x \leftarrow$ random vector from the source 	
\STATE $r \leftarrow$ region $x$ belongs to
\IF{$r$ is already in $regions$}
\STATE $t \leftarrow$ index of $r$ in $regions$
\STATE $centroids[t] \leftarrow centroids[t] + x$
\STATE increment $pointsPerRegion[t]$ of 1
\ELSE
\STATE append $r$ to $regions$
\STATE append $x$ to $centroids$
\STATE append 1 to $pointsPerRegion$ 
\ENDIF
\ENDFOR
\FOR{$k$ in size(centroids)}
\STATE $centroids[k] \leftarrow centroids[k] / pointsPerRegion[k] $
\ENDFOR
\FOR{$numberOfPointsMSE$ times}
\STATE $x$ = random point from the source of information
\STATE $r$ = region of $x$
\STATE $t$ = index of $r$ in $regions$
\STATE $c = centroids[t]$
\STATE $error \leftarrow error + distance(x,c)^2$
\ENDFOR
\STATE $error /= numberOfPointsCentroids$
\end{algorithmic}
\label{al:two}
\end{algorithm}
Note that an approach similar to that of  Algorithm \ref{al:two} is used  is step (i) of the algorithm when estimating the centroid of each region induced by the hyperplane arrangement,  as this evaluation also require numerical integration.

By alternating the optimization step (i) and step (ii) through the MSE approximation approach in Algorithm \ref{al:two}, the algorithm converges to a numerical solution.
Upon multiple random restarts of the algorithm, convergence to multiple local minima is sometimes observed. 
%
%
%
These minimal values arises either from a limitation in the precision of the  numerical integration or by a local minimum in the quantizer configuration. 

\begin{figure}
	\centering
	\includegraphics[width=1\linewidth]{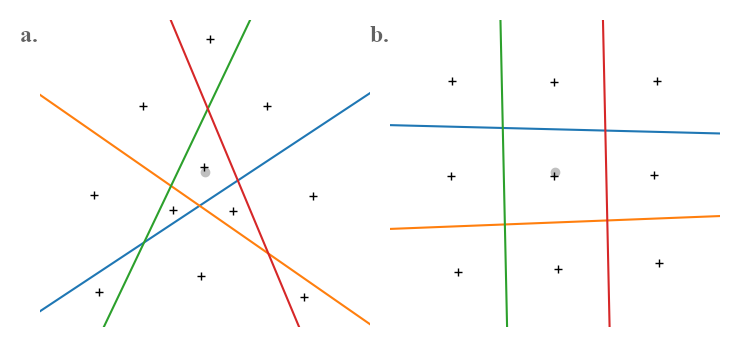}
	\caption{Two approximative local minima of the algorithm in Sec. \ref{sec:Design Algorithm} for the case of a  standard Gaussian distribution. The grey dot represents the center of the distribution.}
	\label{fig:local-minima}
\end{figure}

An example of convergence to multiple  local minima upon multiple random initializations is shown in Fig. \ref{fig:local-minima}:
%
While the arrangement in Fig. \ref{fig:local-minima}.a  has 6 centroids, the one in Fig \ref{fig:local-minima}.b  has 7 but both attain similar performance in the case of a standard Gaussian distribution. 
%
%
%
From a high level perspective, though, one observes that the two configurations are rather distant and the proposed algorithm is not able to converge to the better solution in Fig \ref{fig:local-minima}.b  starting from the configuration in Fig \ref{fig:local-minima}.a.


\begin{remark}
We conjecture that the number of possible matrices $M$ that describe how reconstruction points are obtained from the hyperplane configuration grows only polynomially with $d$ and $\ntq$. 
If this assumption were true, it would be computationally feasible to consider all possible matrices $M$ in order to avoid local minima.
%
Unfortunately, we are currently unable to prove such conjecture.
\end{remark}

\section{Simulation results}
\label{sec:Simulation results}
In this section we present the quantization performance for the quantizer in Sec. \ref{sec:Vector Quantizer Model} for a configuration obtained using the algorithm in  Sec. \ref{sec:Design Algorithm} for the case of 
(i)  standard Gaussian and (ii) unitary uniform distribution.
In both instances, the performance is compared to that attainable using a classic quantizer designed using the classic Max-Lloyd algorithm with the same number of reconstruction points. 

\subsection{Gaussian distribution}

As one could expect, the quantizer obtained performs slightly worse than an optimal quantizer obtained by Max-Lloyd's algorithm with the same number of reconstruction points. However, when comparing the number of hyperplanes used instead of the number of reconstruction points, it performs better than Max-Lloyd's quantizer, as we can see on Fig. \ref{fig: comparison}. The ratio of this algorithm's quantizer's MSE over Max-Lloyd's quantizer's is 0.64 for a configuration with 5 hyperplanes.

A result that may be surprising is that the hyperplanes do not necessarily form the maximum number of regions. Interestingly, they often arrange to make less regions with similar probabilities whether the starting configuration has many regions or not, which turns out to perform relatively well and sometimes better than configurations with more regions.

\subsection{Uniform distribution}

\begin{figure}
	\centering
	\includegraphics[width=1\linewidth]{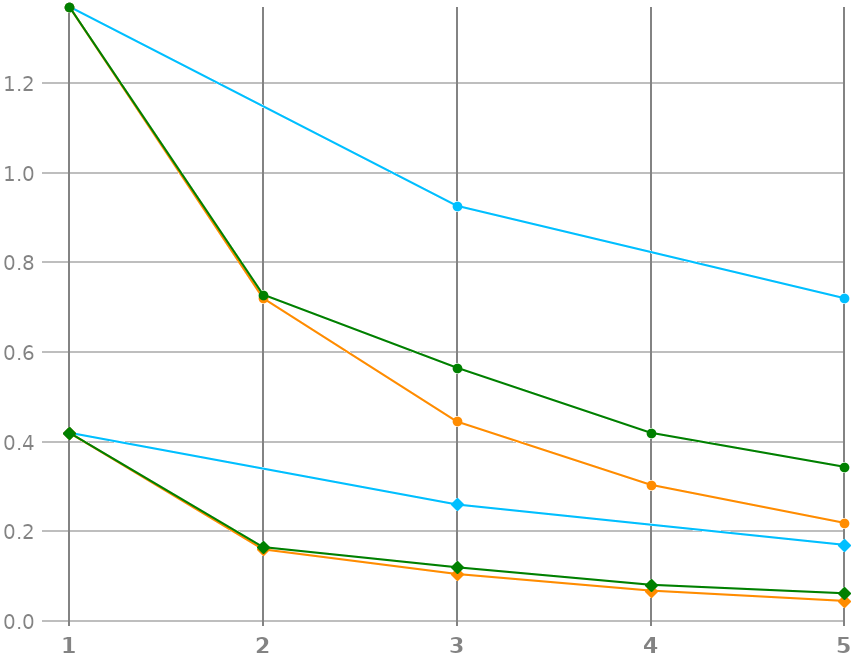}
	\caption{The green curves are the performance of our algorithm depending on the number of comparators. The orange and blue curves are the performances of Max-Lloyd quantizers with the same number of reconstruction points (orange) and the same number of comparators (blue). The curves with circles (the 3 top ones) are for a Gaussian distribution, the ones with squares (the 3 bottom ones) are for a uniform distribution.}
	\label{fig: comparison}
\end{figure}

The results obtained with a uniform distribution are similar to the ones obtained with a Gaussian distribution. Again, the quantizer obtained by the algorithm performs worse than Max-Lloyd's quantizer with the same number of reconstruction points, but better than Max-Lloyd's quantizer with the same number of hyperplanes. This is shown in Fig. \ref{fig: comparison}. The ratio of the MSE of the quantizer obtained over Max-Lloyd's quantizer's is 0.72 for a configuration with 5 hyperplanes.

\begin{figure}
	\centering
	\includegraphics[width=1\linewidth]{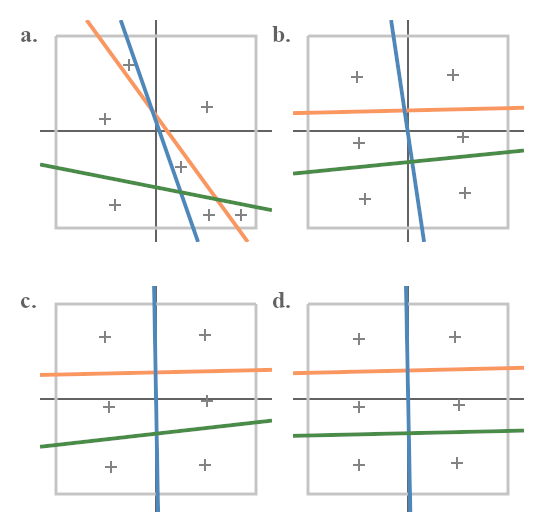}
	\caption{Four steps of optimization with 3 hyperplanes, uniform distribution.  The grey square represents the limit of the distribution support.}
	\label{fig: optimisation-steps-uniform3}
\end{figure}

Because of the square shape of the distribution support, the hyperplanes are even more likely to form rectangular regions than with the Gaussian distribution, as Fig. \ref{fig: optimisation-steps-uniform3} illustrates.

%
%
%
%
%
%
%
%
%
%
%
%
%
%
%
%
%
%

\section{Conclusion}
\label{sec:Conclusion}
In this paper, a novel paradigm for vector quantization is considered. 
In this paradigm, the performance of the quantizer is not limited by the number of reconstruction point as in the classic Max-Lloyd quantizer but rather by the number of comparisons necessary to determine the quantizer output.
In particular, we consider the case in which a vector quantizer is comprised $\ntq$ comparators with receive a linear combination of the quantizer input plus a constant and output the sign of received signal. 
Given a distribution of the quantizer input, $\ntq$ and a distortion measure between source and reconstruction, we consider the problem of optimally determining the linear combination and constant coefficient so that the distortion between source and reconstruction is minimized. 
We propose a first algorithm for this optimization problem and apply such algorithm to the case of mean squared error distortion and Gaussian and uniform iid sources.
In both cases, the performance is compared to that of the Max-Lloyd quantizer.

A number of research directions remain open from this new vector quantizer architecture. In particular, we are investigating the optimal performance attainable in the limit of infinitely long vector quantizer in which the number of available comparators $\ntq$ and bits available to represent the quantizer inputs both grow to infinity at a given constant ratio $\al$. This limit should result in a rather interesting generalization of the classic distortion-rate function.

\bibliographystyle{IEEEtran}
\bibliography{RD_bib}
\end{document}